# Structural and [121]Sb Mössbauer Spectroscopic Investigations of the Antimonide Oxides *RE*MnSbO (*RE* = La, Ce, Pr, Nd, Sm, Gd, Tb) and *RE*ZnSbO (*RE* = La, Ce, Pr)

Inga Schellenberg, Tom Nilges, and Rainer Pöttgen

Institut für Anorganische und Analytische Chemie, Universität Münster, Corrensstrasse 30, D-48149 Münster, Germany

Reprint requests to R. Pöttgen. E-mail: pottgen@uni-muenster.de

((**Heading:** I. Schellenberg *et al.* · Quaternary Antimonide Oxides))

The quaternary antimonide oxides *RE*MnSbO (*RE* = La, Ce, Pr, Nd, Sm, Gd, Tb) and *RE*ZnSbO (*RE* = La, Ce, Pr) were synthesized from the *RE*Sb monoantimonides and MnO, respectively ZnO, in sealed tubes at 1170 K. Single crystals were obtained from NaCl/KCl salt fluxes. The ZrCuSiAs type (space group $P4/nmm$) structures of LaMnSbO ($a$ = 423.95(7), $c$ = 955.5(27) pm, $wR2$ = 0.067, 247 $F^2$), CeMnSbO ($a$ = 420.8(1), $c$ = 950.7(1) pm, $wR2$ = 0.097, 250 $F^2$), SmMnSbO ($a$ = 413.1(1), $c$ = 942.3(1) pm, $wR2$ = 0.068, 330 $F^2$), LaZnSbO ($a$ = 422.67(6), $c$ = 953.8(2) pm, $wR2$ = 0.052, 259 $F^2$), and NdZnSbO ($a$ = 415.9(1), $c$ = 945.4(4) pm, $wR2$ = 0.109, 206 $F^2$) were refined from single crystal X-ray diffractometer data. The structures are composed of covalently bonded $(RE^{3+}O^{2-})^+$ and $(T^{2+}Sb^{3-})^-$ layers with weak ionic-like interlayer interactions. The oxygen and transition metal atoms both have tetrahedral coordination within the layers. [121]Sb

Mössbauer spectra of the $RE$MnSbO and $RE$ZnSbO compounds show single antimony sites with isomer shifts close to −8 mm/s, in agreement with the antimonide character of these samples. PrMnSbO and NdMnSbO show transferred hyperfine fields of 8 T at 4.2 K.

**Introduction**

The quaternary phosphide oxides $RET$PO ($RE$ = rare earth metal; $T$ = transition metal) with $T$ = Mn, Fe, Co, Ni, Zn, Ru, Os [1–12] have intensively been studied in the last 15 years with respect to synthesis conditions and crystal chemistry. Most of these compounds crystallize with the tetragonal ZrCuSiAs type structure [13], space group $P4/nmm$. The structures are composed of two different layers [$RE^{3+}O^{2-}$] and [$T^{2+}P^{3-}$] with essentially covalent $RE$–O and $T$–P bonding within and weak ionic bonding between the layers. Only in the $RE$ZnPO series one observes dimorphism with a rhombohedral high-temperature structure [4, 5, 12]. So far more than 40 $RET$PO compounds are known.

Only recently the physical properties of these materials have been investigated. LaFePO [6] and LaNiPO [7, 11] show a transition to a superconducting state at 3.2 and 4.3 K, respectively. Paramagnetic behavior has been observed for ß-CeZnPO, ß-PrZnPO, GdZnPO [12], CeRuPO and CeOsPO [9]. Among these compounds CeRuPO is a rare example for a ferromagnetic Kondo system.

With the higher homologue arsenic, several series $RET$AsO ($T$ = Mn, Fe, Co, Zn, Ru) [2–4, 10, 14] mostly with the early rare earth elements have been synthesized. Resistivity measurements [10] revealed insulating behavior for LaZnAsO, CeZnAsO, and PrZnAsO, while NdZnAsO is metallic. The cerium, praseodymium, and neodymium compounds show paramagnetic behavior with experimental magnetic moments close to the values of the free $RE^{3+}$ ions. The arsenide oxides show remarkable doping with fluorine, leading to solid solutions $RET$AsO$_{1-x}$F$_x$. This doping induces superconducting transitions at comparatively high tempera-tures, e.g. $T_C = 26$ K in LaFeAsO$_{1-x}$F$_x$ (x = 0.05–0.12) [15]. Most recently an even higher transition temperature of 52 K has been reported for the corresponding praseodymium system [16].

In contrast to the *RE*TPO and *RE*TAsO systems, only few data are available for the systems containing antimony. So far, only the manganese and zinc containing antimonide oxides *RE*MnSbO (*RE* = La–Nd, Sm, Gd) [2, 3] and *RE*ZnSbO (*RE* = La–Nd, Sm) [2, 17, 18] are known. The structures of CeZnSbO [17] and NdMnSbO [3] have been refined from single crystal diffractometer data. Again, CeZnSbO, PrZnSbO, and NdZnSbO show paramagnetism [10, 18] without magnetic ordering down to 3 K.

In the course of our systematic studies of the magnetic and optical properties of these transparent pnictide oxides [5, 12], we have reinvestigated the *RE*TSbO series. Herein we report on single crystal X-ray data of *RE*MnSbO (*RE* = La, Ce, Sm), LaZnSbO and NdZnSbO, and a detailed $^{121}$Sb Mössbauer spectroscopic study of these two series.

**Experimental**

*Synthesis*

Starting materials for the preparation of the antimonide oxides *RE*MnSbO (*RE* = La–Nd, Sm, Gd, Tb) and *RE*ZnSbO (*RE* = La–Nd) were ingots of the rare earth elements (Honeywell, Smart Elements, Chempur, or Kelpin, >99.9 %), Pr$_6$O$_{11}$ (Chempur, >99.9 %), MnO (Sigma Aldrich, > 99.9 %), ZnO (Chempur, >99.5 %), antimony powder (Riedel-de Häen, >99.9 %), zinc drops (Merck, >99.9 %), NaCl (Merck, >99.5 %), and KCl (Chempur, >99.9 %). Small pieces of the rare earth metals were first arc-melted [19] to buttons under an argon atmosphere of ca. 600 mbar. The argon was purified before with titanium sponge (870 K), silica gel and molecular sieves. Three different preparation routes were used.

Most samples were synthesized with the rare earth monoantimonides *RE*Sb as precursors. The latter were obtained by arc-melting from the elements. The *RE*Sb precursors were then mixed with MnO or ZnO powder in 1:1 atomic ratio, finely ground in a mortar and subsequently cold-pressed to pellets of 6 mm diameter. The pellets were placed in small tantalum crucibles and sealed in evacuated silica ampoules. The samples were placed in tube furnaces, heated at a rate of 30 K / h to 1170 K and kept at that temperature for four days. Finally the samples were cooled to room temperature by switching off the power supply. The resulting samples were ground, repressed to pellets and annealed again with the same setup. This procedure was repeated until the samples were single phase. The resulting compounds were obtained in the form of polycrystalline powders.

Alternatively PrZnPO can be synthesized via the precursor compound $PrZn_2$. The latter was synthesized from praseodymium metal and zinc in the ideal 1:2 atomic ratio in a sealed tantalum ampoule. The tantalum tube was placed in a special water-cooled silica sample chamber [20] in an induction furnace (Hüttinger Elektronik, Freiburg, Typ TIG 1.5/300). The mixture was first melted at ca. 1270 K and then annealed at ca. 970 K for 4 hours followed by rapidly cooling by switching off the furnace. The purity of the $PrZn_2$ sample was checked through a Guinier powder pattern. Fine powders of $PrZn_2$, $Pr_6O_{11}$, Zn, and antimony were weighed in the ideal atomic ratio of 5:1:1:11, finely ground, pressed to a small pellet, and annealed in a tantalum crucible under the same conditions as for the $RE$Sb/ZnO mixtures.

Both ceramic preparation routes led only to polycrystalline powders. Well-shaped single crystals of the antimonide oxides were grown in NaCl/KCl salt fluxes. Filings of the rare earth metals, manganese or zinc oxide and powder of antimony were put together in the ideal 1:1:1 atomic ratio and 0.5 g of each mixture was sealed in an evacuated silica ampoule after adding ca. 2 g of an equimolar NaCl/KCl mixture acting as flux medium. The mixtures were heated within one day at 1170 K, kept at that temperature for 6 days and then slowly cooled to 870 K at a rate of 2 K / h followed by cooling to room temperature within one day. The crystals were isolated from the reaction mixtures through repeated extraction of the salt flux with hot demineralised water. All samples are stable in air. Depending on the thickness of the crystals, the $RE$ZnSbO compounds are transparent with a red to dark red color. The *RE*MnSbO crystals show metallic lustre. Powders of these samples are dark grey.

*EDX data*

The single crystals investigated on the diffractometer were studied by energy dispersive analyses of X-rays (EDX) using a Leica 420i scanning electron microscope with $CeO_2$, the rare earth trifluorides, Zn, Mn, and Sb as standards for semiquantitative measurements. The oxygen content of the samples could not be determined reliably (detection limit of the instrument). The experimentally observed ratios *RE* : *T* : Sb of all compounds were within ±2 at-% (standard uncertainty for the analyses at different points) close to the ideal 1 : 1 : 1 ratio. No impurity elements heavier than sodium were detected.

*X-Ray diffraction*

All polycrystalline samples were characterized through X-ray powder diffraction (Guinier technique, imaging plate detector, Fujifilm, BAS–1800 readout system) using Cu$K\alpha_1$ radiation and $\alpha$-quartz ($a$ = 491.30 and $c$ = 540.46 pm) as an internal standard. The tetragonal lattice parameters (Table 1) were refined from the powder data by a least-squares routine. The proper indexing was ensured by intensity calculations [21].

Well-shaped single crystals of *RE*MnSbO (*RE* = La, Ce, Sm), LaZnSbO and NdZnSbO were glued to quartz fibres and their quality was checked by Laue photographs on a Buerger camera using white Mo radiation. Intensity data of CeMnSbO and SmMnSbO were collected at room temperature by use of a four-circle diffractometer (CAD4) with graphite monochromatized Ag$K\alpha$ radiation and a scintillation counter with pulse height discrimination. The scans were taken in the $\omega/2\theta$ mode and empirical absorption corrections were applied on the basis of psi-scan data, accompanied by spherical absorption corrections. The LaZnSbO, NdZnSbO, and LaMnSbO crystals were measured at room temperature by use of a Stoe IPDS-II imaging plate diffractometer in oscillation mode (graphite-monochromatized Mo$K\alpha$ radiation). Numerical absorption corrections were ap-plied to the data sets. All relevant details concerning the data collections and evaluations are listed in Table 2.

*Structure refinements*

The isotypy with the ZrCuSiAs type structure [13] was already evident from the Guinier patterns and all data sets were compatible with space group *P4/nmm*, in agreement with our previous investigations on α-CeZnPO and α-PrZnPO [3]. Consequently, the atomic parameters of α-CeZnPO [3] were taken as starting values and the structures were refined using SHELXL-97 [22] (full-matrix least-squares on $F^2$) with anisotropic atomic displacement parameters for all atoms. As a check for deviations from the ideal composition, all occupancy parameters were refined in separate series of least-squares cycles. Since all sites were fully occupied within one to three standard deviations, the ideal occupancies were assumed again in the final cycles. The final difference Fourier syntheses were flat (Table 2). The positional parameters and interatomic distances are listed in Tables 3 and 4.

Further details on the structure refinements may be obtained from the Fachinformationszentrum Karlsruhe, D-76344 Eggenstein-Leopoldshafen (Germany), by quoting the Registry No's. CSD–419355 (LaMnSbO), CSD–419356 (CeMnSbO), CSD–419357 (SmMnSbO), CSD–419353 (LaZnSbO), and CSD–419354 (NdZnSbO).

*$^{121}$Sb Mössbauer spectroscopy*

A Ba$^{121m}$SnO$_3$ source was used for the Mössbauer spectroscopic experiments. The measurements were carried out in a helium bath cryostat at 4.2 and 77 K. The temperature was controlled by a resistance thermometer (±0.5 K accuracy). The Mössbauer source was kept at room temperature. The samples were enclosed in small PVC containers at a thickness corresponding to about 10 mg Sb/cm$^2$.

**Discussion**

*Crystal chemistry*

The quaternary antimonide oxides *RE*MnSbO (*RE* = La, Ce, Pr, Nd, Sm, Gd, Tb) and *RE*ZnSbO (*RE* = La, Ce, Pr) crystallize with the tetragonal ZrCuSiAs type structure, space group $P4/nmm$. So far, these compounds have been characterized on the basis of X-ray powder data [2, 3, 17, 18] and the structures of CeZnSbO [17] and NdMnSbO [3] have been refined from single crystal diffractometer data. We have now obtained phase pure samples for property measurements and refined also the structures of $RE$MnSbO ($RE$ = La, Ce, Sm), LaZnSbO and NdZnSbO. TbMnSbO is a new member in the $RE$MnSbO series. All these compounds belong to a larger class of pnictide oxides. The crystal chemistry of such materials has been reviewed [23, 24]. Herein we focus only on the structural peculiarities of the tetragonal compounds.

A view of the $RET$SbO structure is shown in Figure 1. These antimonide oxides are composed of two different layers in an AB AB stacking sequence. In view of the transparency of the $RE$ZnSbO compounds, an ionic formula splitting $RE^{3+}T^{2+}Sb^{3-}O^{2-}$ can be assumed. Within the two layer types we observe tetrahedral rare earth coordination for the oxygen atoms and tetrahedral antimony coordination for the transition metals leading to a formulation $[RE^{3+}O^{2-}]^{+}$ $[T^{2+}Sb^{3-}]^{-}$. Similar to the phosphide systems (see the electronic structure calculations of Ce-RuPO [9] and PrZnPO [12]), we can assume strong covalent $RE$–O and $T$–Sb bonding within and weak ionic-like bonding between the layers.

### $^{121}$Sb Mössbauer spectroscopic characterization

The $^{121}$Sb Mössbauer spectra of the antimonide oxides $RE$MnSbO ($RE$ = La, Ce, Pr, Nd, Sm, Gd) and $RE$ZnSbO ($RE$ = La, Ce, Pr) taken at 77 and 4.2 K are presented in Figures 2–5 together with transmission integral fits. The corresponding fitting parameters are listed in Table 5. In agreement with the crystal structures, the spectra could be well reproduced with single antimony sites. Due to the high natural line width of antimony, no quadrupole moment was needed for the fits, however, in view of the non-cubic site symmetry (4$mm$), weak quadrupole splitting is expected.

The isomer shifts (77 K data) range from –7.82 (LaMnSbO) to –8.37 mm/s (PrZnSbO). In view of the ionic formula splitting discussed above the $RET$SbO compounds contains $Sb^{3-}$ pnictide anions. These Zintl anions also occur in the al-kali metal antimonides $A_3$Sb (A = Li, Na, K, Rb) and the experimental isomer shifts are similar, i.e. –7.3 mm/s for Li$_3$Sb [25] and –8.39 mm/s for Rb$_3$Sb [26]. The isomer shifts also compare well with the data for the III–V semiconductors AlSb, GaSb and InSb [27] and the ytterbium based antimonides Yb$T$Sb ($T$ = Ni, Pd, Pt, Cu, Ag, Au) [28].

The observed decrease of the isomer shifts with respect to elemental antimony (–11.6 mm/s) mainly results from an increase of the Sb 5s electron population. This behaviour is discussed for a detailed series of different antimony compounds in references [25] and [27].

At 4.2 K we observe transferred hyperfine fields of ca. 8 T for PrMnSbO and NdMnSbO which result from magnetic ordering of these compounds. All other antimonide oxides studied herein show almost similar patterns at 77 and 4.2 K. Preliminary temperature-dependent susceptibility measurements on the $RE$MnSbO compounds reveal contributions from the rare earth and manganese atoms, leading to complex magnetic ordering. Detailed investigations are in progress.

*Acknowledgements*

This work was financially supported by the Deutsche Forschungsgemeinschaft. We thank Dipl.-Ing. U. Ch. Rodewald and B. Heying for collecting the single-crystal X-ray diffraction data and Dipl.-Chem. F. M. Schappacher for experimental help for the Mössbauer spectroscopic experiments.

Table 1. Lattice parameters of the tetragonal antimonide oxides *RE*MnSbO and *RE*ZnSbO at 293 K.

| Compound | $a$ (pm) | $c$ (pm) | $V$ (nm$^3$) | Reference |
|----------|----------|----------|--------------|-----------|
| LaMnSbO | 423.95(7) | 955.5(2) | 0.1717 | this work |
|  | 424.2(1) | 955.7(2) | 0.1720 | [3] |
| CeMnSbO | 420.8(1) | 950.7(1) | 0.1683 | this work |
|  | 421.8(1) | 951.7(2) | 0.1693 | [3] |
| PrMnSbO | 418.8(1) | 947.2(3) | 0.1661 | this work |
|  | 418.7(1) | 946.0(1) | 0.1658 | [3] |
| NdMnSbO | 416.6(1) | 947.1(3) | 0.1644 | this work |
|  | 416.5(1) | 946.2(2) | 0.1641 | [3] |
| SmMnSbO | 413.1(1) | 942.3(1) | 0.1608 | this work |
|  | 413.5(1) | 941.8(2) | 0.1610 | [3] |
| GdMnSbO | 410.0(1) | 942.1(2) | 0.1584 | this work |
|  | 409.0(1) | 941.0(1) | 0.1574 | [3] |
| TbMnSbO | 408.3(1) | 939.2(6) | 0.1566 | this work |
| LaZnSbO | 422.67(6) | 953.8(2) | 0.1704 | this work |
|  | 422.62(2) | 953.77(6) | 0.1704 | [17] |
|  | 422.604(7) | 953.691(24) | 0.1703 | [10] |
| CeZnSbO | 419.9(1) | 948.7(2) | 0.1673 | this work |
|  | 419.76(4) | 947.4(1) | 0.1669 | [17] |
|  | 419.66(2) | 947.96(4) | 0.1669 | [10] |
| PrZnSbO | 418.79(8) | 946.7(5) | 0.1660 | this work |
|  | 417.63(4) | 945.1(1) | 0.1648 | [17] |
| NdZnSbO | 415.9(1) | 945.4(4) | 0.1635 | this work |
|  | 415.81(2) | 944.95(5) | 0.1634 | [17] |
|  | 415.78(2) | 944.33(5) | 0.1632 | [10] |
| SmZnSbO | 412.80(2) | 940.16(6) | 0.1602 | [17] |

Table 2. Crystal data and structure refinement for *RET*SbO, ZrCuSiAs-type, space group *P*4/*nmm*, *Z* = 2.

| | LaMnSbO | CeMnSbO | SmMnSbO | LaZnSbO | NdZnSbO |
|---|---|---|---|---|---|
| Empirical formula | LaMnSbO | CeMnSbO | SmMnSbO | LaZnSbO | NdZnSbO |
| Molar mass, g/mol | 331.60 | 332.81 | 343.04 | 342.03 | 347.36 |
| Unit cell dimensions | Table 1 | Table 1 | Table 1 | Table 1 | Table 1 |
| Calculated density, g/cm$^3$ | 6.41 | 6.57 | 7.09 | 6.67 | 7.05 |
| Crystal size, µm | 20 × 40 × 40 | 10 × 25 × 35 | 20 × 40 × 40 | 20 × 50 × 90 | 10 × 40 × 40 |
| Detector distance, mm | 60 | – | – | 80 | 80 |
| Exposure time, min | 5 | – | – | 7 | 12 |
| $\omega$ range; increment, deg | 0–180; 1.0 | – | – | 0–180; 1.0 | 0–180; 1.0 |
| Integr. param. A, B, EMS | 13.5; 3.5; 0.012 | – | – | 13.5; 3.5; 0.012 | 13.5; 3.5; 0.012 |
| Transm. ratio (max/min) | 3.59 | 1.64 | 1.06 | 7.19 | 1.29 |
| Absorption coefficient, mm$^{-1}$ | 23.4 | 13.0 | 15.9 | 26.9 | 30.8 |
| $F(000)$, e | 282 | 284 | 292 | 292 | 298 |
| $\theta$ range, deg | 4 to 34 | 3 to 27 | 3 to 30 | 4 to 35 | 2 to 32 |
| Range in *hkl* | –5 / 6, ±6, ±15 | ±6, ±6, ±15 | ±7, ±7, ±16 | ±6, ±6, –15 / 12 | –6 / 5, ±6, ±14 |
| Total no. reflections | 2394 | 2683 | 3758 | 2523 | 1969 |
| Independent reflections | 247 ($R_{int}$ = 0.087) | 250 ($R_{int}$ = 0.283) | 330 ($R_{int}$ = 0.090) | 259 ($R_{int}$ = 0.070) | 206 ($R_{int}$ = 0.111) |
| Reflections with $I \geq 2\sigma(I)$ | 187 ($R_\sigma$ = 0.037) | 164 ($R_\sigma$ = 0.111) | 273 ($R_\sigma$ = 0.035) | 208 ($R_\sigma$ = 0.030) | 189 ($R_\sigma$ = 0.040) |
| Data/parameters | 247 / 12 | 250 / 12 | 330 / 12 | 259 / 12 | 206 / 12 |
| Goodness-of-fit on $F^2$ | 1.005 | 1.009 | 1.133 | 1.077 | 1.031 |
| Final *R* indices [$I \geq 2\sigma(I)$] | *R*1 = 0.026 | *R*1 = 0.043 | *R*1 = 0.033 | *R*1 = 0.021 | *R*1 = 0.044 |
| | *wR*2 = 0.065 | *wR*2 = 0.085 | *wR*2 = 0.064 | *wR*2 = 0.050 | *wR*2 = 0.107 |
| *R* indices (all data) | *R*1 = 0.035 | *R*1 = 0.088 | *R*1 = 0.045 | *R*1 = 0.029 | *R*1 = 0.047 |
| | *wR*2 = 0.067 | *wR*2 = 0.097 | *wR*2 = 0.068 | *wR*2 = 0.052 | *wR*2 = 0.109 |
| Extinction coefficient | 0.028(3) | 0.004(5) | 0.034(4) | 0.031(3) | 0.033(6) |
| Largest diff. peak and hole, e Å$^{-3}$ | 2.57 / –4.07 | 2.57 / –3.54 | 1.92 / –5.80 | 2.76 / –3.00 | 3.13 / –3.95 |

Table 3. Atomic positions and isotropic displacement parameters (pm$^2$) of LaMnSbO, CeMnSbO, SmMnSbO, LaZnSbO, and NdZnSbO. $U_{eq}$ is defined as one third of the trace of the orthogonalized $U_{ij}$ tensor.

| Atom | Wyckoff position | $x$ | $y$ | $z$ | $U_{eq}$ |
|------|------------------|-----|-----|-----|----------|
| **LaMnSbO** | | | | | |
| La | 2$c$ | 1/4 | 1/4 | 0.11951(7) | 48(2) |
| Mn | 2$b$ | 3/4 | 1/4 | 1/2 | 83(4) |
| Sb | 2$c$ | 1/4 | 1/4 | 0.68112(8) | 72(2) |
| O | 2$a$ | 3/4 | 1/4 | 0 | 32(13) |
| **CeMnSbO** | | | | | |
| Ce | 2$c$ | 1/4 | 1/4 | 0.11850(13) | 64(4) |
| Mn | 2$b$ | 3/4 | 1/4 | 1/2 | 102(7) |
| Sb | 2$c$ | 1/4 | 1/4 | 0.68328(17) | 84(4) |
| O | 2$a$ | 3/4 | 1/4 | 0 | 31(26) |
| **SmMnSbO** | | | | | |
| Sm | 2$c$ | 1/4 | 1/4 | 0.11421(6) | 60(2) |
| Mn | 2$b$ | 3/4 | 1/4 | 1/2 | 94(3) |
| Sb | 2$c$ | 1/4 | 1/4 | 0.68793(8) | 83(2) |
| O | 2$a$ | 3/4 | 1/4 | 0 | 70(12) |
| **LaZnSbO** | | | | | |
| La | 2$c$ | 1/4 | 1/4 | 0.12088(5) | 53(2) |
| Zn | 2$b$ | 3/4 | 1/4 | 1/2 | 128(3) |
| Sb | 2$c$ | 1/4 | 1/4 | 0.67982(7) | 85(2) |
| O | 2$a$ | 3/4 | 1/4 | 0 | 35(10) |
| **NdZnSbO** | | | | | |
| Nd | 2$c$ | 1/4 | 1/4 | 0.11714(7) | 94(4) |
| Zn | 2$b$ | 3/4 | 1/4 | 1/2 | 172(6) |
| Sb | 2$c$ | 1/4 | 1/4 | 0.68477(10) | 127(4) |
| O | 2$a$ | 3/4 | 1/4 | 0 | 78(19) |

Table 4. Interatomic distances, calculated with the powder lattice parameters of LaMnSbO, CeMnSbO, SmMnSbO, LaZnSbO, and NdZnSbO. Standard deviations are all equal or smaller than 0.2 pm.

| | | | LaMnSbO | CeMnSbO | SmMnSbO | LaZnSbO | NdZnSbO |
|--|--|--|---------|---------|---------|---------|---------|
| *RE*: | 4 | O | 240.8 | 238.7 | 232.9 | 240.7 | 235.6 |
| | 4 | Sb | 355.2 | 352.2 | 346.5 | 354.2 | 348.7 |
| *T*: | 4 | Sb | 273.7 | 273.2 | 272.1 | 272.2 | 271.6 |
| Sb: | 4 | *T* | 273.7 | 273.2 | 272.1 | 272.2 | 271.6 |
| | 4 | *RE* | 355.2 | 352.2 | 346.5 | 354.2 | 348.7 |
| O: | 4 | *RE* | 240.8 | 238.7 | 232.9 | 240.7 | 235.6 |

Table 5. Fitting parameters of $^{121}$Sb Mössbauer spectroscopic measurements of different quaternary antimonide oxides. Numbers in parentheses represent the statistical errors in the last digit. ($\delta$), isomer shift; ($\Gamma$), experimental line width.

| Compound | T / K | $\delta$ / mm·s$^{-1}$ | $\Gamma$ / mm·s$^{-1}$ |
|---|---|---|---|
| LaMnSbO | 77 | -7.82(1) | 2.78(2) |
| CeMnSbO | 77 | -7.89(1) | 2.95(9) |
| | 4.2 | -7.90(3) | 3.2(2) |
| PrMnSbO | 77 | -8.03(2) | 3.49(4) |
| | 4.2 | -8.12(4) | 2.7(2) |
| NdMnSbO | 77 | -7.99(1) | 2.78(9) |
| | 4.2 | -7.93(3) | 2.8(2) |
| SmMnSbO | 77 | -7.94(1) | 2.82(4) |
| | 4.2 | -8.02(3) | 3.1(2) |
| GdMnSbO | 77 | -8.11(2) | 3.1(1) |
| | 4.2 | -8.32(3) | 3.4(3) |
| LaZnSbO | 77 | -8.15(2) | 2.92(8) |
| CeZnSbO | 77 | -8.15(1) | 2.66(6) |
| | 4.2 | -8.33(3) | 3.1(2) |
| PrZnSbO | 77 | -8.37(2) | 2.78(4) |

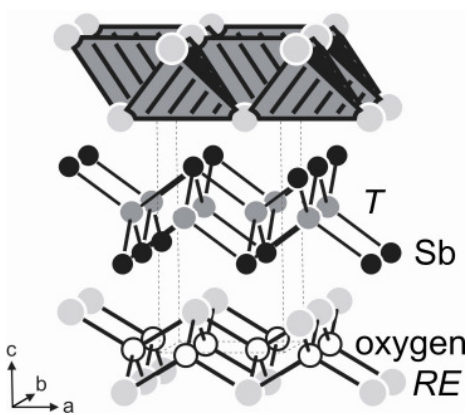

Fig. 1. The crystal structure of *RET*SbO. The different layers of condensed O*RE*$_4$ and *T*Sb$_4$ tetrahedra are emphasized.

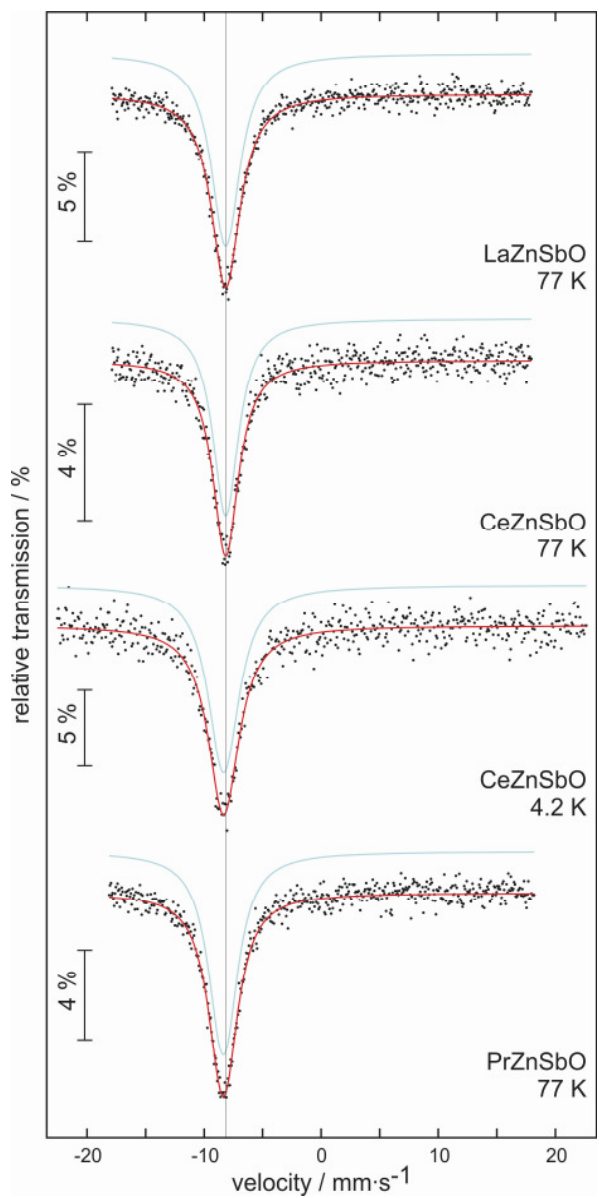

Fig. 2. Experimental and simulated [121]Sb Mössbauer spectra of LaZnSbO, CeZnSbO, and PrZnSbO at different temperatures. The vertical line serves as a guide for the eye.

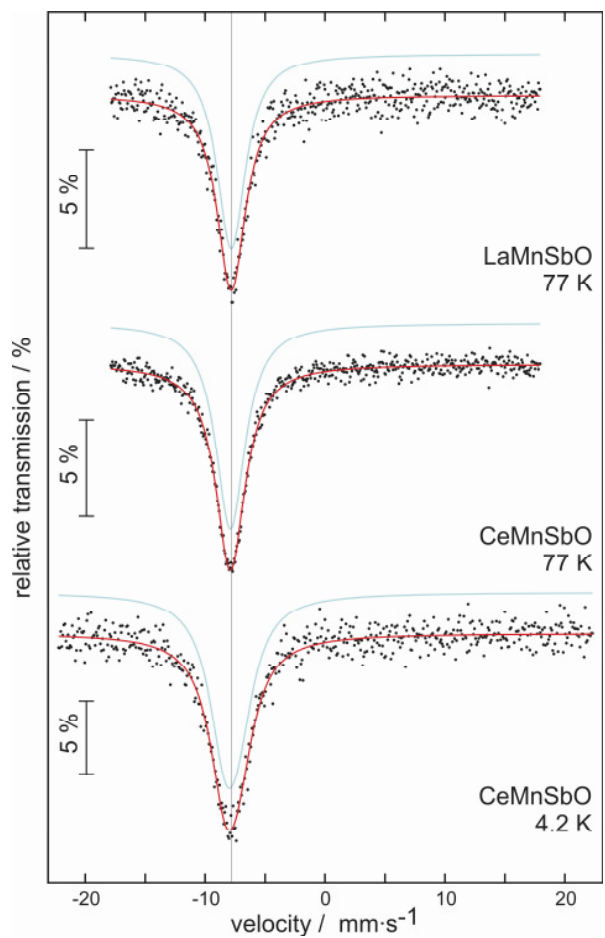

Fig. 3. Experimental and simulated $^{121}$Sb Mössbauer spectra of LaMnSbO and CeMnSbO at different temperatures. The vertical line serves as a guide for the eye.

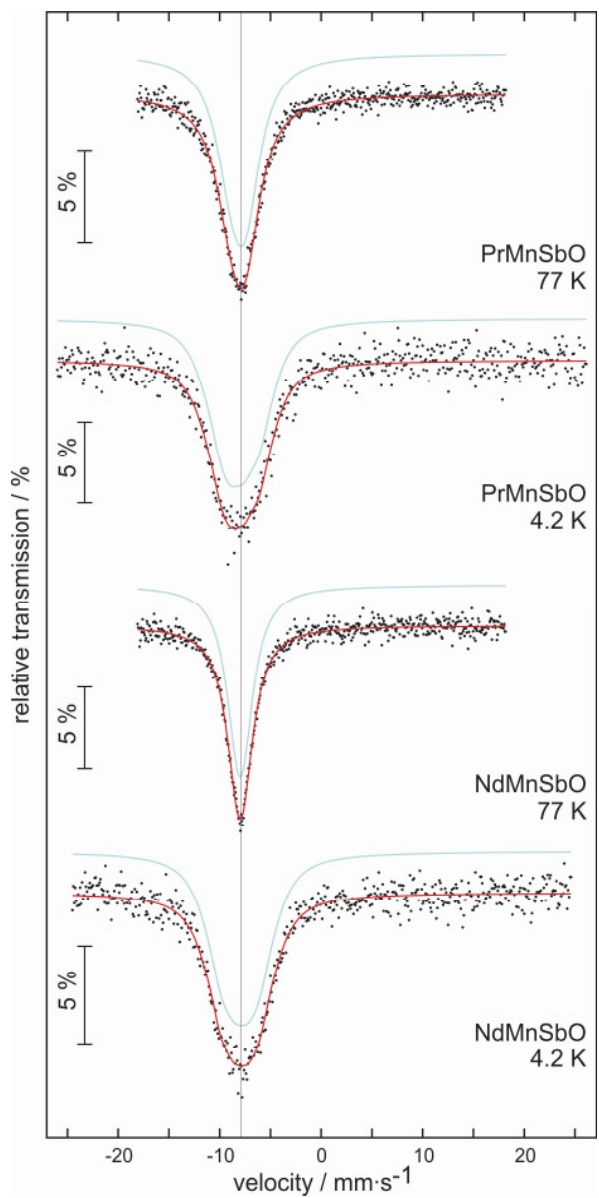

Fig. 4. Experimental and simulated $^{121}$Sb Mössbauer spectra of PrMnSbO and NdMnSbO at 77 and 4.2 K. The vertical line serves as a guide for the eye.

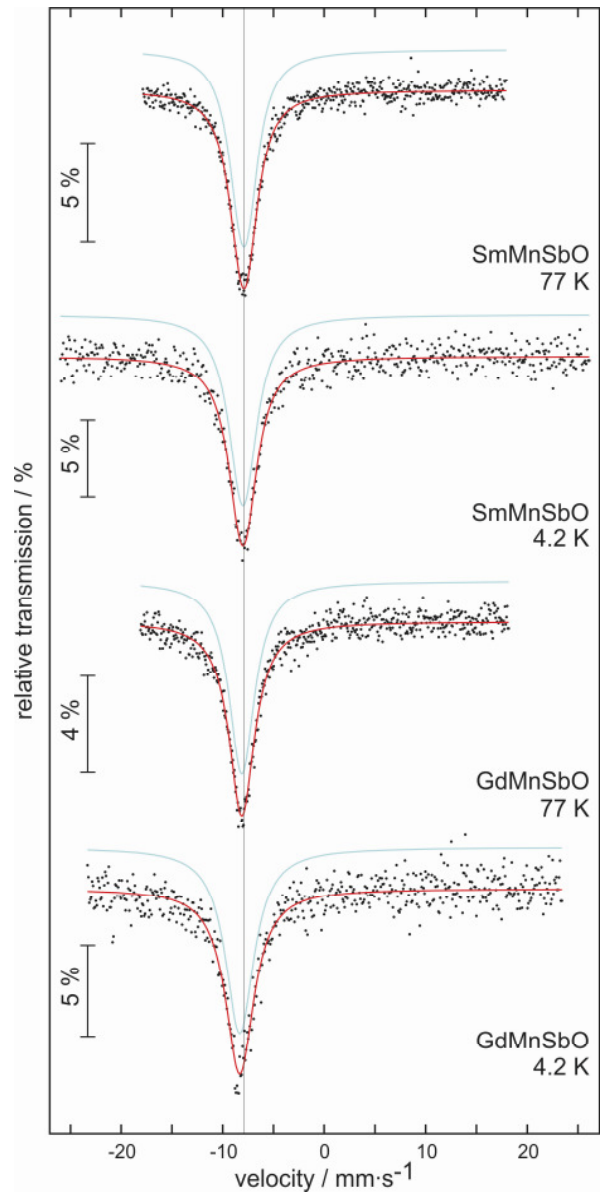

Fig. 5. Experimental and simulated [121]Sb Mössbauer spectra of SmMnSbO and GdMnSbO at 77 and 4.2 K. The vertical line serves as a guide for the eye.